\documentclass[pra,twocolumn, showpacs, showkeys, secnumarabic, aps, amsmath, amssymb, nofootinbib, superscriptaddress, longbibliography, floatfix, table-of-contents, dblfloatfix]{revtex4-2}

\usepackage[utf8]{inputenc}
\usepackage[pdftex]{graphicx}
\usepackage{mathrsfs}
\usepackage[colorlinks, breaklinks, urlcolor={blue}, linkcolor={red}, citecolor={blue}]{hyperref}
\usepackage{array}
\usepackage{amsmath}
\usepackage{type1cm}
\usepackage{lettrine}
\usepackage[english]{babel}
\usepackage{lmodern}
\usepackage{microtype}
\usepackage{booktabs}
\usepackage[T1]{fontenc}
\usepackage[boxed, vlined]{algorithm2e}
\usepackage{caption}
\usepackage{braket}
\usepackage{xcolor}
\usepackage{orcidlink}

\begin{document}

\title{Certified Quantumness via Single-Shot Temporal Measurements}

\author{Md~Manirul Ali \orcidlink{0000-0002-5076-7619}}
\email{manirul@citchennai.net}
\affiliation{Centre for Quantum Science and Technology, Chennai Institute of Technology, Chennai 600069, India}

\author{Sovik Roy \orcidlink{0000-0003-4334-341X}}
\email{s.roy2.tmsl@ticollege.org}
\affiliation{Department of Mathematics, Techno Main Salt Lake (Engg. Colg.), Techno India Group, EM 4/1, Sector V, Salt Lake, Kolkata  700091, India}

\begin{abstract}
Bell-Kochen-Specker theorem states that a non-contextual hidden-variable theory cannot completely reproduce the predictions of quantum mechanics. Asher Peres gave a remarkably simple proof of quantum contextuality in a four-dimensional Hilbert space of two spin-1/2 particles. Peres's argument is enormously simpler than that of Kochen and Specker. Peres contextuality demonstrates a logical contradiction between quantum mechanics and the noncontextual hidden variable models by showing an inconsistency when assigning noncontextual definite values to a certain set of quantum observables. In this work, we present a similar proof in time with a temporal version of the Peres-like argument. In analogy with the two-particle version of Peres's argument in the context of spin measurements at two different locations, we examine here single-particle spin measurements at two different times $t=t_1$ and $t=t_2$. We adopt three classical assumptions for time-separated measurements, which are demonstrated to conflict with quantum mechanical predictions. Consequently, we provide a non-probabilistic proof of certified quantumness in time, without relying on inequalities, demonstrating that our approach can certify the quantumness of a device through single-shot, time-separated measurements. Our results can be experimentally verified with the present quantum technology.
\end{abstract}

\maketitle

In 1935, Einstein, Podolski, and Rosen claimed that the description of physical reality provided by quantum mechanics is {\it incomplete},
and a more complete realistic theory with predetermined definite values of physical quantities is possible. Realistic hidden-variable models
seek to provide a ``complete specification'' of the state of a quantum system so that the individual measurement values of a dynamical
variable is predetermined by some hidden-variables \cite{Bell66,Mermin93}. In 1964, Bell showed \cite{Bell64} that a local hidden-variables
model cannot completely reproduce the measurement statistics or statistical correlations predicted by quantum mechanics, such types of hidden
variable models have been experimentally ruled out through the violation of Bell's inequalities \cite{BellExpt1,BellExpt2,BellExpt3,BellExpt4,BellExpt5,BellExpt6}. Inspired by Bell's theorem, Leggett and Garg \cite{Leggett85} introduced a temporal analogue of Bell inequalities, expressed in terms of time-separated measurement statistics or temporal correlation functions-specifically, the Leggett-Garg inequalities (LGI) test correlations within a single system measured at different times. LGI are based on two classical assumptions: ($a_1$) macrorealism, which posits that macroscopic systems always exist in definite states with well-defined pre-existing values, and ($a_2$) noninvasive measurability, which asserts that these values can be measured without disturbing the system. However, quantum systems often violate these assumptions \cite{LGExpt1,LGExpt2,LGExpt3,LGExpt4,LGExpt5,LGExpt6}. The violation of LGI by quantum systems suggests either the breakdown of a realistic description or the impossibility of noninvasive measurement. Such violations are evidence of nonclassical behavior (or quantumness) \cite{RevNori14,MMA14,MMA17}, and have also been used to certify
quantum randomness \cite{DHome24}.

Another ``no-hidden-variables'' theorem known as Bell-Kochen-Specker (BKS) theorem, demonstrates the inconsistency between
quantum mechanics and noncontextual hidden-variables models \cite{Bell66,Kochen67} by showing a contradiction when assigning
noncontextual predetermined values to a certain set of quantum observables. Local hidden variable theories are a special type of
noncontextual hidden variable theories where the context is provided through spatial separation \cite{Das13,Cabello08}. The original
proof \cite{Kochen67} of quantum contextuality by Kochen and Specker involves a complex structure using 117 vectors
in a three dimensional space. Subsequently, new versions of the BKS theorem have been proposed
\cite{Peres90,Mermin90,Kernaghan,Penrose94,Cabello96,Cabello98,SimonPRL,DHome,Cabello08} that are simple to
understand. Quantum contextuality is shown to provide advantages in implementing quantum information processing tasks
\cite{Adv01,Adv02,Adv03,Adv04,Adv05}. Several experimental tests of quantum contextuality have also been carried out
\cite{Yuji06,Kirchmair09,Amselem09,Zeilinger11,PRX13}. Noncontextual hidden variable theories assign measurement values
to an observable which is predetermined regardless of which other compatible observable is measured jointly with that observable.
Quantum contextuality implies that the results of measuring an observable depends on the set of other co-measurable compatible
measurements that are being performed jointly or sequentially. A remarkably simple proof of quantum contextuality was provided
by Peres in a four-dimensional Hilbert space of two spin-1/2 particles \cite{Peres90}. Peres's argument is enormously simpler
than that of Kochen and Specker. Peres quantum contextuality demonstrates a logical contradiction between quantum mechanics
and the noncontextual hidden variable models by showing an inconsistency when assigning noncontextual definite values to a
certain set of quantum observables.

In this work, we present a similar proof in time with a temporal version of the Peres-like contextuality
argument \cite{Peres90}. We adopt here Leggett-Garg's classical realism ($a_1$) and noninvasive measurability ($a_2$)
hypothesis along with another classical assumption ($a_3$) that any set of physical observables of a classical system can have
simultaneous predetermined measurement values. The above assumptions will be shown to conflict with quantum mechanical
predictions, and our argument can potentially be used to certify the quantumness of a device using single-shot time-separated
measurements. Please note that we present here a non-probabilistic proof of certified quantumness in time without inequalities,
where the time dynamics of the spin observables are regulated by Heisenberg equations of motion. In fact, we provide here a
temporal version of the Peres-like argument analogous to the GHZ version \cite{GHZ89,GHZ90,MerminAm90} of Bell's theorem
without inequalities, which do not require any statistical correlation functions associated to the probabilities of measurement
outcomes for many repeated joint measurements or any statistical inequalities. Like GHZ and Peres arguments, here in this
work we used eigenstates of three mutually commuting Hermitian operators (time-dependent) to build a logical contradiction
that is sharp and extreme in the sense that for some particular measurement, $+1$ result is expected whereas quantum mechanics
results in $-1$ instead.

The notion of ``noncontextuality'' adopted by Peres is as follows. Let $A$ and $B$ be two mutually
commuting observables. According to some realistic hidden-variable model, let $v(A)$ be the individual measurement outcome
of $A$ that is predetermined by a hidden variable $\lambda$. Let $v(B)$ and $v(AB)$ be the individual measurement values
of the observables $B$ and $AB$ respectively, which are also predetermined by the same hidden variable. Then the
``noncontextuality'' of that hidden variable model is characterized by the following feature \cite{Peres90}
\begin{eqnarray}
\label{ab}
v(AB) = v(A) v(B).
\end{eqnarray}
The above assumptions are shown to be inconsistent with quantum mechanical prediction, which implies that value assignment is
not possible or the value-assignment must depend on which context the observable appears in. This phenomenon is known as quantum
contextuality. Peres considered two spatially separated spin-1/2 particles prepared in a singlet state
\begin{eqnarray}
|\Psi\rangle = \frac{1}{\sqrt{2}} \left(|\uparrow\rangle_1 |\downarrow\rangle_2
- |\downarrow\rangle_1 |\uparrow\rangle_2 \right),
\label{Sing}
\end{eqnarray}
where $|\uparrow\rangle$ and $|\downarrow\rangle$ represent the eigenstates of spin operator $\sigma_z$
along $z$-axis with eigenvalues $+1$ and $-1$ respectively. One can then consider measurements of three
mutually commuting Hermitian operators in this two-qubit Hilbert space as
\begin{eqnarray}
\label{x1x2}
\sigma_x^1~\sigma_x^2, \\
\label{y1y2}
\sigma_y^1~\sigma_y^2, \\
\label{z1z2}
\sigma_z^1~\sigma_z^2.
\end{eqnarray}
These three operators can have simultaneous eigenstates and eigenvalues as they all commute with each other. One can verify that the
singlet state $|\Psi\rangle$ is an eigenstate of the three product operators $\sigma_x^1 \sigma_x^2$, $\sigma_y^1 \sigma_y^2$,
and $\sigma_z^1 \sigma_z^2$ with a common eigenvalue $-1$. Moreover, the Pauli spin operators obey the relations $\sigma_z^1=-i\sigma_x^1\sigma_y^1$, and $\sigma_z^2=i\sigma_y^2\sigma_x^2$. Then according to quantum mechanics,
we can write down the eigenvalue equations for the three mutually commuting operators given by Eqs.~(\ref{x1x2}), (\ref{y1y2}),
and (\ref{z1z2}) as follows
\begin{eqnarray}
\label{x1x2e}
\sigma_x^1~\sigma_x^2~|\Psi\rangle &=& (-1)~|\Psi\rangle, \\
\label{y1y2e}
\sigma_y^1~\sigma_y^2~|\Psi\rangle &=& (-1)~|\Psi\rangle, \\
\label{z1z2e}
\sigma_x^1~\sigma_y^2~\sigma_y^1~\sigma_x^2~|\Psi\rangle &=& (-1)~|\Psi\rangle,
\end{eqnarray}
where we have used $\sigma_z^1~\sigma_z^2$$=\sigma_x^1~\sigma_y^1~\sigma_y^2~\sigma_x^2$
$=\sigma_x^1~\sigma_y^2~\sigma_y^1~\sigma_x^2$. The quantum measurement results
(\ref{x1x2e}-\ref{z1z2e}) cannot be interpreted by a noncontextual hidden-variables model.
Let one try to assign dichotomic values to the outcomes of each local Pauli spin measurements $\sigma_x^1$,
$\sigma_x^2$, $\sigma_y^1$, $\sigma_y^2$ as the corresponding eigenvalues of the operators $v_r^i=\pm 1$,
$r \in {x,y}$ and $i \in {1,2}$. These value assignments are independent on which context or on which sequence
the observables appear in. These values are predetermined by some noncontextual hidden-variables and they
must satisfy the relations
\begin{eqnarray}
\label{value1}
v_x^1~v_x^2 &=& -1, \\
\label{value2}
v_y^1~v_y^2 &=& -1, \\
\label{value3}
v_x^1~v_y^2~v_y^1~v_x^2 &=& -1,
\end{eqnarray}
in order to reproduce the predictions of quantum mechanics (\ref{x1x2e}-\ref{z1z2e}). Hence for the two spin-1/2 particles
prepared in a state $|\Psi\rangle$, quantum mechanics predicts that the product of the measurement outcomes for the three
commuting observables $\sigma_x^1 \sigma_x^2$, $\sigma_y^1 \sigma_y^2$, and $\sigma_x^1 \sigma_y^2 \sigma_y^1 \sigma_x^2$
should be $-1$. But according to noncontextual hidden-variables scheme, each local spin measurement along a particular axis occurs
twice in the overall set of three product measurements, and according to the left hand sides of Eqs.~(\ref{value1}), (\ref{value2})
and (\ref{value3}), the product of the predetermined measurement values should be $(v_x^1)^2 (v_x^2)^2 (v_y^1)^2 (v_y^2)^2 = +1$.
This is in sharp contradiction with the product of the right hand sides approved by the results of quantum mechanics. Hence, if the local
spin components of each spin-1/2 particles is pre-assigned before the measurement, we end up with a logical contradiction with respect
to quantum measurements on two-particle entangled state $|\Psi\rangle$. Peres contextuality demonstrates the contextual nature of
quantum mechanics that violates our intuition of classical noncontextual realism.

Next, we present a temporal version of the Peres-like argument where the measurements are separated in time.
Specifically, instead of considering spatially separated spin measurements in four-dimensional Hilbert space, here we consider
time separated spin measurements in two-dimensional Hilbert space of a single spin-1/2 system.  Analogous to the two-particle
version of Peres's contextuality argument, here we consider single-particle spin measurements at two different times $t=t_1$
and $t=t_2$ with $t_2>t_1$. We then choose following three joint measurement operators that corresponds to sequential
measurements
\begin{eqnarray}
\label{xy}
&\sigma_x(t_2)~\sigma_y(t_1), \\
\label{yx}
&\sigma_y(t_2)~\sigma_x(t_1), \\
\label{zz}
&\sigma_z(t_2)~\sigma_z(t_1).
\end{eqnarray}

The measurement outcomes of an observable in quantum mechanics are determined by the eigenvalues of the
Hermitian operator that is related to that observable. Under a classical hidden-variables scheme,
one can attempt to assign values to the individual spin components at different times as the corresponding
eigenvalues of the operators associated to the time-dependent spin observables. This attempt of assigning
values to individual measurement outcomes predetermined by the hidden variable will fail as is shown
below. Let us consider a single spin-1/2 system evolving under a particular Hamiltonian
$H=\frac{\hbar}{2} \omega \sigma_z$. The time evolution of the Pauli spin observables are governed by the
Heisenberg equations of motion
\begin{eqnarray}
\label{sxh}
&\frac{d}{dt} \sigma_x(t) = \frac{1}{i\hbar} \left[ \sigma_x(t),  H \right], \\
\label{syh}
&\frac{d}{dt} \sigma_y(t) = \frac{1}{i\hbar} \left[ \sigma_y(t),  H \right], \\
\label{szh}
&\frac{d}{dt} \sigma_z(t) = \frac{1}{i\hbar} \left[ \sigma_z(t),  H \right],
\end{eqnarray}
with the Heisenberg picture operators
\begin{eqnarray}
\label{sxt}
&\sigma_x(t) = e^{i H t / \hbar} ~ \sigma_x ~ e^{ - i H t / \hbar}, \\
\label{syt}
&\sigma_y(t) = e^{i H t / \hbar} ~ \sigma_y ~ e^{ - i H t / \hbar}, \\
\label{szt}
&\sigma_z(t) = e^{i H t / \hbar} ~ \sigma_z ~ e^{ - i H t / \hbar}.
\end{eqnarray}

For our specific Hamiltonian $H=\frac{\hbar}{2} \omega \sigma_z$, the exact time evolution
of the spin operators are given by
\begin{eqnarray}
\label{sxt2}
\sigma_x(t) &=& \sigma_x \cos(\omega t) - \sigma_y \sin(\omega t), \\
\label{syt2}
\sigma_y(t) &=& \sigma_y \cos(\omega t) + \sigma_x \sin(\omega t), \\
\label{szt2}
\sigma_z(t) &=& \sigma_z
\end{eqnarray}
where $\sigma_{x}$, $\sigma_{y}$, and $\sigma_{z}$ are standard Pauli spin operators that represent measurements
of spin along $x$, $y$, and $z$ direction respectively. We now consider two instants of time $t_1$ and $t_2$, so
that $\omega t_1=0$ and $\omega t_2=\pi/2$. Then according to the time evolution equations (\ref{sxt2}),
(\ref{syt2}) and (\ref{szt2}), the three operators given by Eqs.~(\ref{xy}), (\ref{yx}) and (\ref{zz}) become
Hermitian operators for these choices of times $t_1$ and $t_2$. Under that situation
all three joint operators (\ref{xy}-\ref{zz}) become mutually commuting with each other. Hence, the three
operators can have simultaneous eigenstates and eigenvalues as they all commute with each other. One can
verify that a given state $|\Phi\rangle=|\uparrow\rangle$ is an eigenstate
of the operator $\sigma_x(t_2) \sigma_y(t_1)$ with eigenvalue $-1$. Simultaneously, the state $|\Phi\rangle$ is
also an eigenstate of the operators $\sigma_y(t_2) \sigma_x(t_1)$ and $\sigma_z(t_2) \sigma_z(t_1)$ with a
common eigenvalue $+1$. Now the Pauli spin operators for this single spin-1/2 system obey the relations
$\sigma_x(t) \sigma_y(t)$$=i\sigma_z(t)$, and $\sigma_y(t) \sigma_x(t)$$=-i\sigma_z(t)$ so that at any
particular instant of time $t$ the commutation relation $[\sigma_x(t),\sigma_y(t)]=2i\sigma_z(t)$ is satisfied.
According to quantum mechanics, we can then write down the eigenvalue equations for the three mutually commuting
Hermitian operators given by Eqs.~(\ref{xy}), (\ref{yx}) and (\ref{zz}) as follows
\begin{eqnarray}
\label{xye}
\sigma_x(t_2)~\sigma_y(t_1)~|\Phi\rangle &=& (-1)~|\Phi\rangle, \\
\label{yxe}
\sigma_y(t_2)~\sigma_x(t_1)~|\Phi\rangle &=& (+1)~|\Phi\rangle, \\
\label{zze}
\sigma_x(t_2)~\sigma_y(t_2)~\sigma_y(t_1)~\sigma_x(t_1)~|\Phi\rangle &=& (+1)~|\Phi\rangle.
\end{eqnarray}
We now pursue the idea of classical hidden variables$-$the outcomes of measuring observables have already been
specified prior to measurement. We adopt here three classical assumptions ($a_1$) Classical realism: classical systems always
exist in definite states with well-defined pre-existing values, ($a_2$) Noninvasive measurability: those pre-existing values can be
measured without disturbing the system, and ($a_3$) Simultaneous measurability: any set of physical observables of a
classical system can have simultaneous predetermined measurement values. Under these classical assumptions, the measurement values
of the individual time-dependent spin observables $\sigma_x(t_1)$, $\sigma_y(t_1)$, and $\sigma_x(t_2)$, $\sigma_y(t_2)$
are predetermined by some classical hidden-variables. Analogous to the two-particle version of Peres's contextuality
argument involving spin measurements at two different locations, we examine single-particle spin measurements at two distinct
times, $t=t_1$ and $t=t_2$, where the context is defined by the time separation. According to the above classical assumptions,
we assign the values of the observables irrespective of the sequence in which the observables appear. Whether we perform any
measurement or not, the $x$-component of spin $\sigma_x$ has predetermined values $m_x^1$, $m_x^2$, at times $t_1$, and
$t_2$ respectively. Similarly, the predetermined measurement values for the operators $\sigma_y(t_1)$, $\sigma_y(t_2)$ are
assigned as $m_y^1$, and $m_y^2$ respectively. These values are independent of whether $\sigma_x$ is measured before
or after $\sigma_y$. Those fixed values are predetermined by some classical hidden variables, and they are independent of the
sequence at which the sets of time-separated spin measurements are performed. Now for our suitably chosen two different times
$t_1$ and $t_2$, and for a given state $|\Phi\rangle$, the three mutually commuting Hermitian operators have a set of simultaneous
eigenvalues $-1$, $+1$, and $+1$ respectively. We have taken $\omega t_1=0$ and $\omega t_2=\pi/2$. Then according to the
requirement of quantum mechanics, consistent with the eigenvalue equations Eqs.~(\ref{xye}-\ref{zze}) we need to satisfy the
following relations
\begin{eqnarray}
\label{xyv}
m_x^2~m_y^1 &=& -1, \\
\label{yxv}
m_y^2~m_x^1 &=& +1, \\
\label{zzv}
m_x^2~m_y^2~m_y^1~m_x^1 &=& +1.
\end{eqnarray}

Note that in the above relations, the individual value of any one of the quantities $m_x^1$, $m_x^2$, $m_y^1$, $m_y^2$ is
the same irrespective of the equation in which it occurs. For example, the value of $m_x^2$ is the same in Eqs.~(\ref{xyv})
and (\ref{zzv}), which is predetermined by some classical hidden variables. It is implicitly assumed classically (\ref{zzv})
that the spin observables $\sigma_x$ and $\sigma_y$ can have simultaneous measurement outcomes $m_x^1$ and $m_y^1$
at time $t_1$, and $m_x^2$ and $m_y^2$ at time $t_2$. Now, multiplying the left-hand sides of equations (\ref{xyv}-\ref{zzv})
yields $(m_x^1)^2(m_x^2)^2(m_y^1)^2(m_y^2)^2=+1$, which is in contradiction with the product of right-hand sides.
The product of the right-hand sides of equations (\ref{xyv}-\ref{zzv}) gives $-1$. This implies that quantum mechanics is
inconsistent with classical value assignments for time-separated measurements. This result establishes a Peres-like
argument in time for a single qubit. Our result can potentially be used to certify the quantumness of a device
using single-shot time-separated measurements. It is important to note that our analysis is state-independent because
Eqs.~(\ref{xye}), (\ref{yxe}), and (\ref{zze}) are satisfied by any state for this single spin-$1/2$ system. The non-probabilistic
proof of certified quantumness in time without inequalities, discovered in this work can be experimentally verified with the
present NMR technique \cite{Mahesh13,Kavita22}, superconducting quantum technology \cite{SCQ}, or with trapped ion
qubits \cite{Kirchmair09}.

\vskip 0.5cm

{\bf Declaration of competing interest} The authors declare that they have no known competing financial interests or
personal relationships that could have appeared to influence the work reported in this paper.

\vskip 0.5cm

{\bf Data availability statement} All data that support the findings of this study are included within the article
(and any supplementary files).

\end{document}